\documentclass[%
 reprint,
 amsmath,amssymb,
 aps,
]{revtex4-2}
\usepackage{xcolor}

\usepackage{graphicx}
\usepackage{dcolumn}
\usepackage{bm}
\usepackage[utf8]{inputenc}
\usepackage{comment}

\begin{document}
\preprint{APS/123-QED}

\title{Optical Darboux Transformer}

\author{Auro M. Perego}%
 \email{a.perego1@aston.ac.uk}
\affiliation{%
 Aston Institute of Photonic Technologies, Aston University, Birmingham, B4 7ET, United Kingdom
}%

\date{\today}

\begin{abstract}
The Optical Darboux Transformer for solitons is introduced as a photonic device which performs the Darboux transformation directly in the optical domain. This enables two major advances for optical signal processing based on the nonlinear Fourier transform: (i) the multiplexing of solitonic waveforms corresponding to different discrete eigenvalues of the Zakharov-Shabat system, and (ii) the  selective filtering of an arbitrary number of individual solitons too. The Optical Darboux Transformer can be built using existing commercially available photonic technology components and constitutes a universal tool for signal processing, optical communications, optical rogue waves generation, and waveform shaping and control in the nonlinear Fourier domain.
\end{abstract}

\maketitle

\begin{figure}[!h]
    \centering
\includegraphics[width=10.5cm]{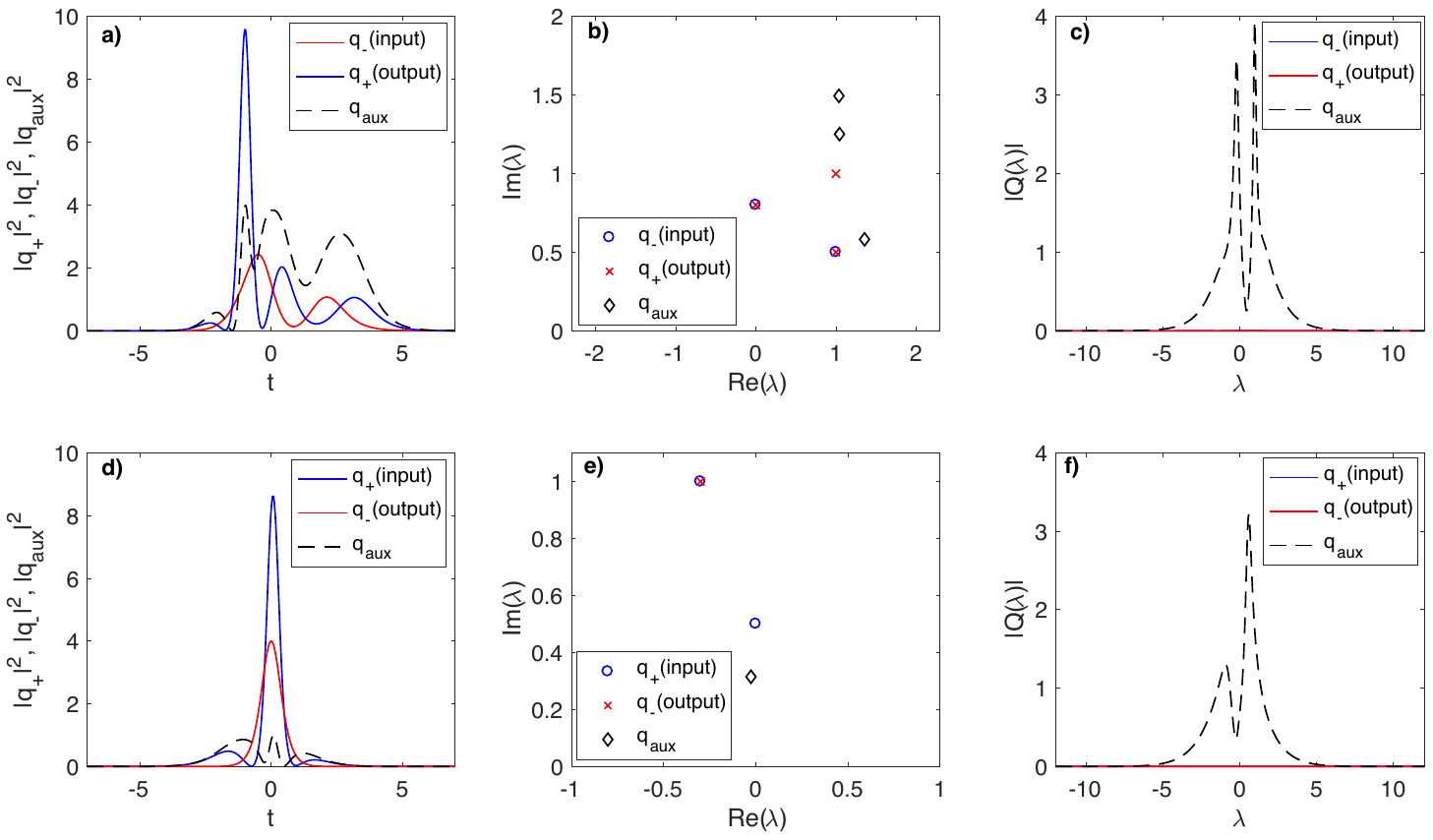}
    \caption{\emph{Single eigenvalue addition and filtering.}
In the multiplexer mode the waveforms, discrete, and continuous spectrum are shown in panels a)-c) for input, output and auxiliary signal (see legend): the input signal $q_-$ features spectrum $\{\lambda_1=1+0.5i,\lambda_2=0.8i\}$ with spectral coefficients $\{b_1=5,b_2=0.5\}$ while the output $q_+$ features the same eignevalues plus the extra eigenvalue $\lambda_3=1+i$ with $b_3=-1$. The filter mode is illustrated in panels d)-f): from the input signal $q_+$ with spectrum $\{\lambda_1=-0.3+i,\lambda_2=0.5i\}$ and $\{b_1=1,b_2=0.8\}$ the eigenvalue $\lambda_2$ is removed.}
    \label{fig:2}
\end{figure}

\begin{figure}[tbp]
    \centering
    \includegraphics[width=10.5cm]{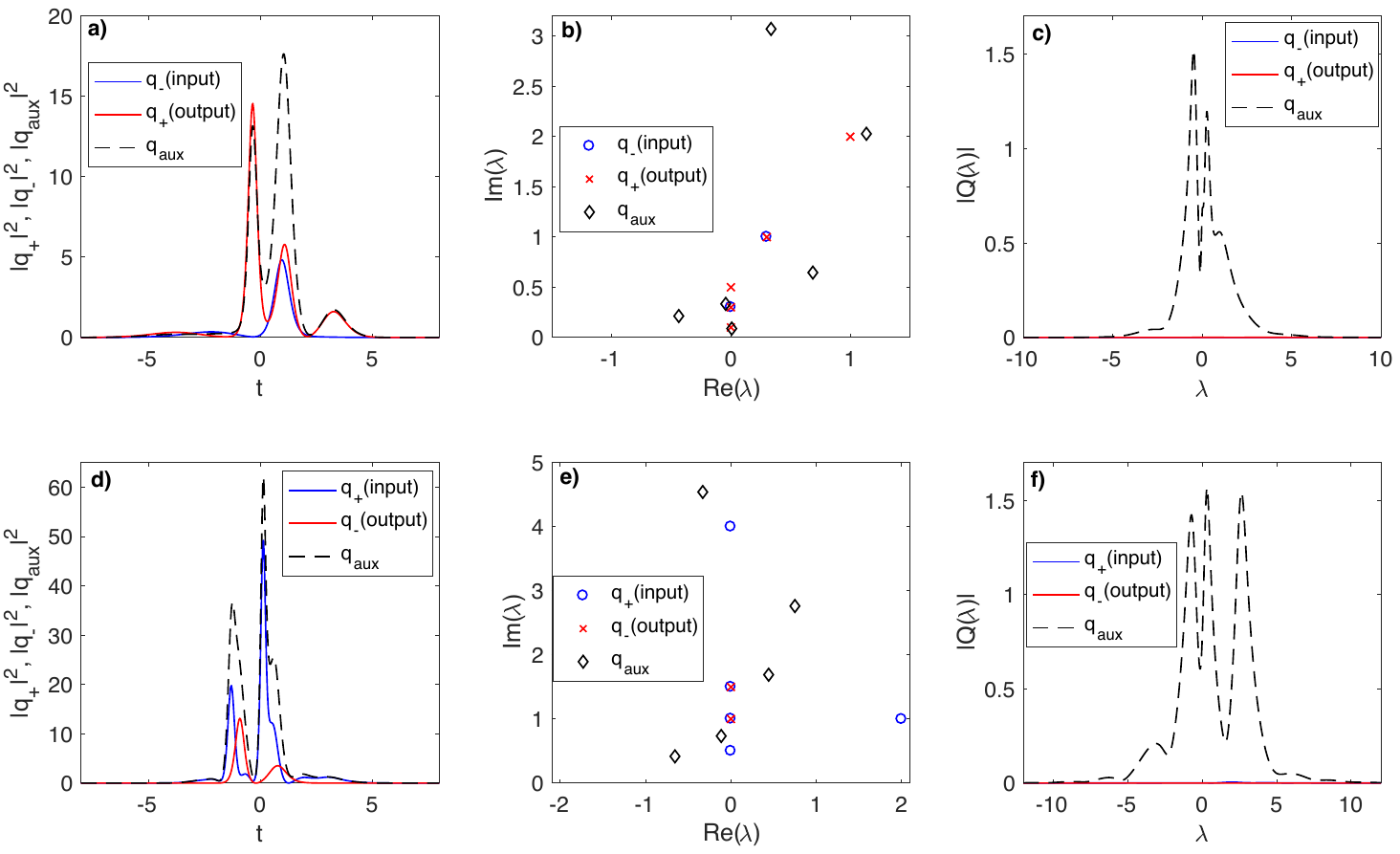}
    \caption{\emph{Multiple eigenvalues addition and filtering.} Multiple eigenvalues addition is shown in panels a)-c) for the input signal $q_-$ having spectrum $\{\lambda_1=0.3+i,\lambda_2=0.3i\}$ with $\{b_1=5, b_2=0.5\}$ where the ODT adds to it the following additional eigenvalues $\{\lambda_3=1+2i,\lambda_4=0.5i,\lambda_5=0.1i\}$ with $\{b_3=-1, b_4=2, b_5=0.1\}$. Multiple eigenvalues filtering is shown in panels d)-f).
    Input signal $q_+$ has spectrum $\{\lambda_1=i,\lambda_2=1.5i,\lambda_3=0.5i,\lambda_4=4i,\lambda_5=2+i\}$ with $\{b_1=1, b_2=0.2, b_3=-2, b_4=3, b_5=1\}$, and eigenvalues $\lambda_3$, $\lambda_4$, and $\lambda_5$ have been removed.}
    \label{fig:3}
\end{figure}

\begin{figure}[!h]
    \centering
    \includegraphics[width=10.5cm]{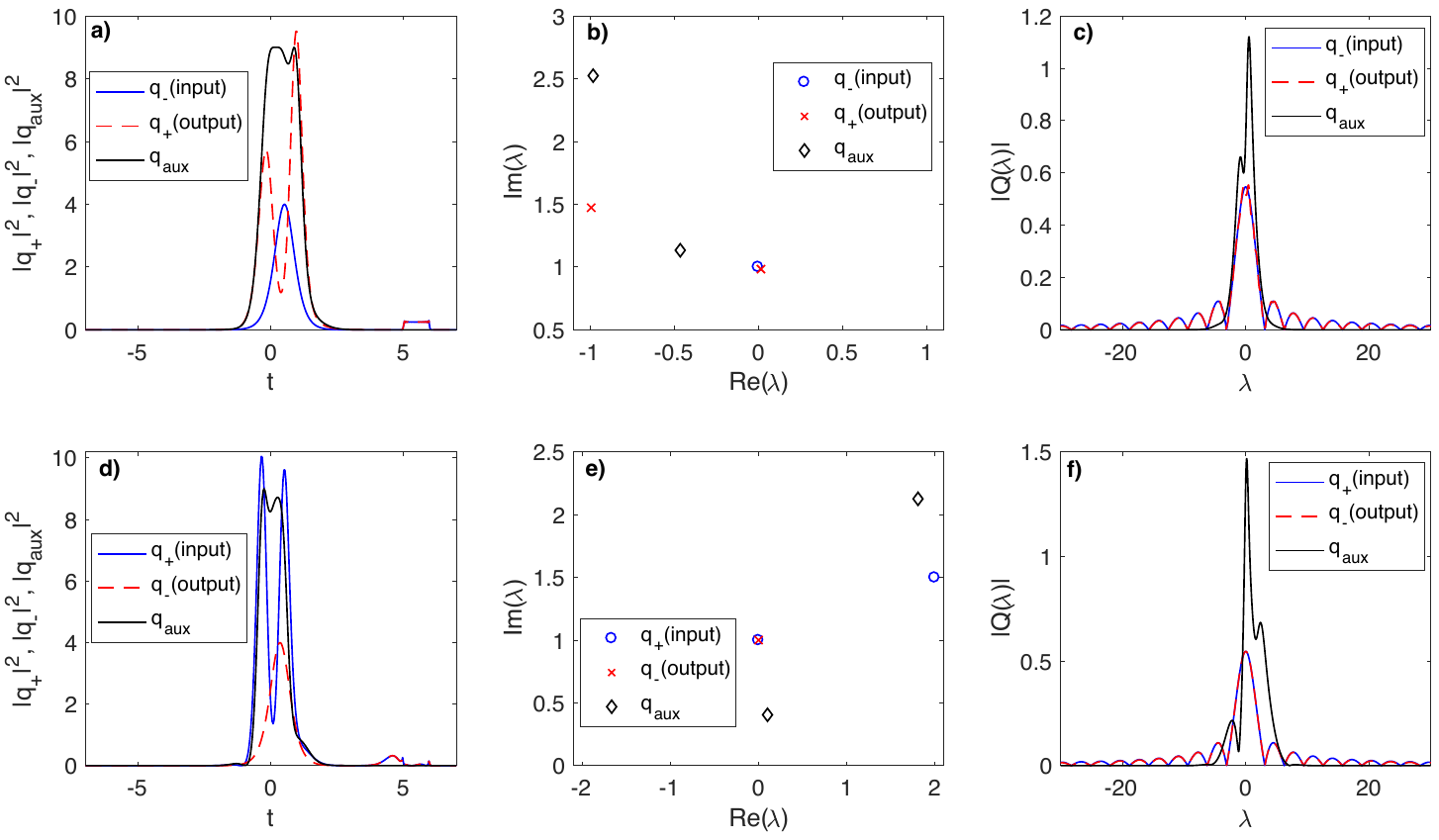}
    \caption{\emph{Eigenvalues multiplexing and filtering in presence of continuous spectrum}. In panels a)-c) the multiplexing case is shown for input signal $q_-$ featuring discrete spectrum $\{\lambda_1=i\}$ with $\{b_1=3\}$, and continuous spectrum $Q(\lambda)= A/(i\lambda)e^{-2i\lambda t_d}\left(1-\frac{\Delta}{i\lambda}\text{cot}(\Delta T)\right)^{-1}$ with $\sqrt{|A|^2+\lambda^2}$, $A=0.5$, $t_d=6$, $T=1$. The ODT output provides signal $q_+$ featuring the additional eigenvalue $\{\lambda_2=-1+1.5i\}$ with $\{b_2=-4\}$. In d)-f) a scenario is shown where  input signal $q_+$, featuring discrete spectrum $\{\lambda_1=i,\lambda_2=2+1.5i\}$ with $\{b_1=2,b_2=-1\}$, and continuous spectrum $Q_f(\lambda)=\prod_{i=1}^2\frac{\lambda-\lambda_i}{\lambda-\lambda_i^*}Q_m(\lambda)$, is transformed into the new signal $q_-$ differing by eigenvalue $\{\lambda_2=2+1.5i\}$ with $\{b_2=-1\}$.}
    \label{fig:4}
\end{figure}

The concept of optical soliton has been proposed 50 years ago as a temporally localised light pulse having hyperbolic secant shape and being invariant upon spatial propagation. This pulse is the analytical solution to the nonlinear Schr{\"o}dinger equation (NLSE) describing the evolution of the electric field slowly varying envelope along a single mode optical fiber\cite{Hasegawa}. Most importantly, Zakharov and Shabat in their seminal paper of 1972 \cite{ZS}, have shown how to obtain bright soliton solutions to the NLSE using the mathematical technique called inverse scattering transform (IST)\cite{NovikovBook,Ablowitz2}. Given a particular complex waveform $q(t)$, function of time $t$, from the solutions of the Zakharov-Shabat system -- a system of two coupled linear equations which depend on $q$ -- it is possible to define a nonlinear spectrum of the waveform. This nonlinear spectrum consists in general of a set of discrete points (denoted by eigenvalues $\lambda_i$) in the complex half-plane -- each of them corresponding to a soliton -- and of a continuous part $Q(\lambda)$ (being $\lambda$ a spectral parameter performing a role analogous to frequency) which is associated to dispersive waves. The latter tends to the classical Fourier spectrum in the low signal power limit \cite{Mansoor}. The nonlinear spectrum provides a set of orthogonal nonlinear eigenmodes which are invariant upon propagation and constitute a complete basis for the description of a given signal \cite{Ablowitz}. Solving the Zakharov-Shabat system for signal $q(t)$ and finding the associated nonlinear spectrum, consists in performing the Nonlinear Fourier Transform (NFT) of $q(t)$.
\newline
In the decades following the pioneering contributions by Hasegawa and Tappert and by Zakharov and Shabat, substantial research on optical solitons has taken place both from a mathematical \cite{NovikovBook,AblowitzBook} and from a physics and engineering perspective too \cite{HasegawaBook,MollenauerBook}.
Optical solitons have been used in commercially deployed optical communication systems and their concept has been generalised to encompass robust localised solutions existing in dissipative systems too -- non integrable by the IST -- thanks to a balance between nonlinearity, dispersion gain/injection and losses \cite{AkhBook}. In this broader sense dissipative optical solitons are a cornerstones of modern laser mode-locking \cite{Akhmediev}, and a workhorse for optical frequency combs generation in nonlinear driven optical resonators \cite{Kippenberg}.
In the recent years, building on the concept of eigenvalue communications proposed by Hasegawa and Nyu \cite{HasegawaNyu,MARUTA}, a novel paradigm for soliton based optical communication systems in fibers has emerged. This approach aims at encoding information in the discrete spectrum eigenvalues (solitons and their phases) and in the continuous spectrum too\cite{Turitsyn:17,NIS,Mansoor,Aref:18,Son,Gaiarin:18}. Further to that, NFT has been suggested as well as a powerful tool for the analysis of optical signals generated by dissipative optical devices like mode-locked lasers\cite{NFTlaser,Chekhovskoy} and optical resonators\cite{Turitsyn:20,TURmr1} too.
This is enabled by the fact that the nonlinear spectrum -- the union of the continuous and discrete parts -- of a given function provides a complete basis for the description of signals with vanishing boundary conditions at $t\rightarrow\pm\infty$ \cite{Ablowitz,Mansoor}. Hence one can project a signal on the nonlinear spectrum basis obtaining its decomposition in terms of solitons and continuous spectrum and then compute new NFTs after the signal has evolved, monitoring in this fashion the nonlinear spectral changes associated also with consequent creation and annihilation of spectral components. This mathematical basis featuring discrete eigenvalues proves advantageous for following the evolution of coherent structures embedded inside a complex waveform. Indeed via its discrete spectral points, NFT enables the description of localised structures using individual eigenvalues, hence condensing the key information in a reduced number of degrees of freedom (pulse peak, temporal position, velocity and phase) compared to a multifrequency classical Fourier spectrum.
\newline
Despite these recent advances, NFT applications in photonics remain very much a technology in its infancy, suffering from limitations ranging from poor algorithmic performances for computing the nonlinear spectrum to the lack of optical domain nonlinear signal processing techniques. For instance, apparently simple problems such as (i) how to optically combine different waveforms containing multiple solitons so that the novel waveform contains exactly the solitons present into the two original signals, or (ii) how to optically eject or filter individual solitons from a signal without irreversibly destroying the whole waveform, remained unsolved so far. This is rooted in the qualitatively different nonlinear physics of solitons compared to the linear principles which inspire the customary approach to signal processing. For instance, a traditional optical filter is not able to discriminate between a soliton and the continuous spectrum; nor simply coherently combining a soliton with a signal waveform using for instance an optical coupler will add a discrete eigenvalue to the nonlinear spectrum of the signal. 
\newline
To address these fundamental limitations we present here a novel device based on the optical implementation of a mathematical operation called the Darboux transformation (DT), which enables optical multiplexing of solitons to an existing waveform -- including in presence of continuous spectrum -- and selective removal of optical solitons from a signal too, without undesired effects on the waveform spectral properties.
Mathematically the DT is  transformation which takes as an input a solution to the NLSE and a new discrete spectrum eigenvalue, providing as an output a new solution to the NLSE having the same nonlinear spectrum as the input one plus the new discrete eigenvalue \cite{MatveevBook,Aref:18,Mansoor3}. The DT offers a simple and iterative procedure to build multisoliton solutions to the NLSE and its generalised forms. For this reason DT has been widely used in NFT based optical communication systems to digitally generate multisoliton signals both in single\cite{Mansoor3,Aref:18,Bulow,aref,Son} and dual-polarisation regimes\cite{Gaiarin:18,DaRos:19}; and for finding rogue waves solutions too\cite{rogue}. However, in the examples reported so far its implementation does not occur in optical domain. The DT is iteratively calculated digitally till a waveform $q(t)$ featuring the desired nonlinear spectrum is obtained. Then the function  $q(t)$ is transferred from the electrical to the optical domain by means of a wave form generator and an optical modulator which suitably modifies a laser beam. This poses a fundamental limitation to solitonic waveform signal processing and namely the necessity of completely passing from the optical to the electrical domain and then back to optical one every time one needs to manipulate a signal e.g. for multiplexing, demultiplexing, regeneration, and noise filtering. To the best of our knowledge no optical implementation of the DT has been ever proposed so far. The goal of this work is to introduce a method for performing the DT optically, and to practically answer the question: how can one selectively add and remove discrete eigenvalues from an optical signal without undesired effects on the spectral properties of the signal itself?\newline
First, we need to briefly introduce the known DT formalism for the nonlinear Schr{\"o}dinger equation (NLSE)\cite{aref}.
Let us consider a given signal $q(z,t)$ whose propagation along spatial coordinate $z$ is ruled by the focussing normalised NLSE:
\begin{eqnarray}
\frac{\partial q}{\partial z}=i\frac{\partial^2q}{\partial t^2}+2i|q|^2q.
\end{eqnarray}
The normalised amplitude, spatial and temporal coordinates are related to physical fiber and electric field parameters by the following relations: $q=E/\sqrt{P}$, $t=\tau/T_0$, and $z=-Z/\mathcal{L}$; where $E$ is the electric field slowly varying envelope in units $\sqrt{\text{W}}$, $P=|\beta_2|/(\gamma T_0^2)$, $\mathcal{L}=2T_0^2/|\beta_2|$, $\beta_2$ is the group velocity dispersion in ps$^2$km$^{-1}$, $\gamma$ is the nonlinearity coefficient in km$^{-1}$W$^{-1}$, $\tau$ is a the temporal coordinate in a reference frame co-moving with the pulse, $Z$ is the spatial coordinate along the fiber in physical units, and $T_0$ is a free normalization parameter.
Starting from the knowledge of $q(t)$ it is possible to obtain a new solution to the NLSE $\tilde{q}$ which features the same discrete spectrum as $q$ plus a new soliton described by the discrete eigenvalue $\lambda_0$. If $q$ had non vanishing continuous spectrum $Q_c(\lambda)$ then $\tilde{q}$ has continuous spectrum given by $\tilde{Q}_c(\lambda)=\frac{\lambda-\lambda_0^*}{\lambda-\lambda_0}Q_c(\lambda)$. The new solution is defined by

\begin{eqnarray}\label{DT}
\tilde{q}(t)=q(t)+2i(\lambda_0^*-\lambda_0)\frac{v_2^*v_1}{|v_1|^2+|v_2|^2}
\end{eqnarray}
which is the DT.
$v_1(t)$ and $v_2(t)$ are the two components of the vector $v=(v_1,v_2)^T$ which is a solution to the Zakharov-Shabat system for the signal $q(t)$. The vector 
\begin{eqnarray}
\tilde{v}=(\lambda_0\bold{I}_2-\bold{G}_0)v
\end{eqnarray}
is instead the solution to the Zakharov-Shabat system for signal $\tilde{q}$; being
$\bold{I}_2$ the $2\times2$ identity matrix and $\bold{G}_0=\bold{\Theta}\bold{M}_0\bold{\Theta}^{-1}$ with $\bold{M}_0=\text{diag}(\lambda_0,\lambda_0^*)$ and
\begin{eqnarray}
\bold{\Theta}=\begin{pmatrix} 
v_1 & v_2^* \\
v_2 & -v_1^*
\end{pmatrix}.
\end{eqnarray}
\newline
Now we generalise the DT to the case when a signal $\tilde{q}$ is generated starting from the vacuum solution $q=0$ (or from a solution having no discrete spectrum but instead continuum spectrum $Q_c(\lambda)$), and features $n$ discrete eigenvalues $\{\lambda_1,..,\lambda_n\}$ which are iteratively added to the spectrum of $q$. This is naturally achieved by iterating $n$ times the procedure highlighted in Eq.\ref{DT}.
We have \begin{eqnarray}\label{DT2}
\tilde{q}(t)=q(t)+\sum_{i=1}^n \mathcal{D}_i
\end{eqnarray}
where
$\mathcal{D}_i=2i(\lambda_i^*-\lambda_i)\frac{v_2^{(i)*}v_1^{(i)}}{|v_1^{(i)}|^2+|v_2^{(i)}|^2}$ with the precaution that for every $\lambda_i$ added in the iterative process, the functions $v_{1,2}^{(k)}$ associated to the remaining -- yet to be added -- eigenvalues $\lambda_k$ ($k=i+1,...,n$) have to be updated as follows:
\begin{eqnarray}\label{vk}
\begin{pmatrix} 
\tilde{v}_1^{(k)} \\
\tilde{v}_2^{(k)}
\end{pmatrix}=(\lambda_k\bold{I}_2-\bold{G}_{oi})...(\lambda_k\bold{I}_2-\bold{G}_{01})\begin{pmatrix} 
A^{(k)}e^{-i\lambda_kt} \\
B^{(k)}e^{i\lambda_kt}
\end{pmatrix}
\end{eqnarray}
being the matrix $\bold{G}_{0i}$ computed as a function of $v_{1,2}^{(i)}$ evaluated after $i-1$ iterations of the DT. It can be furthermore shown that if one wants to generate a signal featuring $n$ eigenvalues $\{\lambda_1,..\lambda_n\}$ corresponding to solitons having spectral amplitudes determined by the spectral coefficients $\{b_1,...b_n\}$, then the initialisation of the coefficients should be as follows:
$A^{(i)}=1$ and $B^{(i)}=b_i$\cite{aref}.
Note also that if $q$ is not the vacuum solution but it has a continuous spectrum $Q_c(\lambda)$, then one should initialise the iterative DT algorithm with signal $q'$ (instead of $q$), with the spectrum $Q_c'(\lambda)=Q_c(\lambda)\prod_{i=1}^n\frac{\lambda-\lambda_i}{\lambda-\lambda_i^*}$ if one wants to have $\tilde{q}$ with continuous spectrum $Q_c$\cite{aref}.
\newline
\begin{figure}[tbp]
    \centering
    \includegraphics[width=8.5cm]{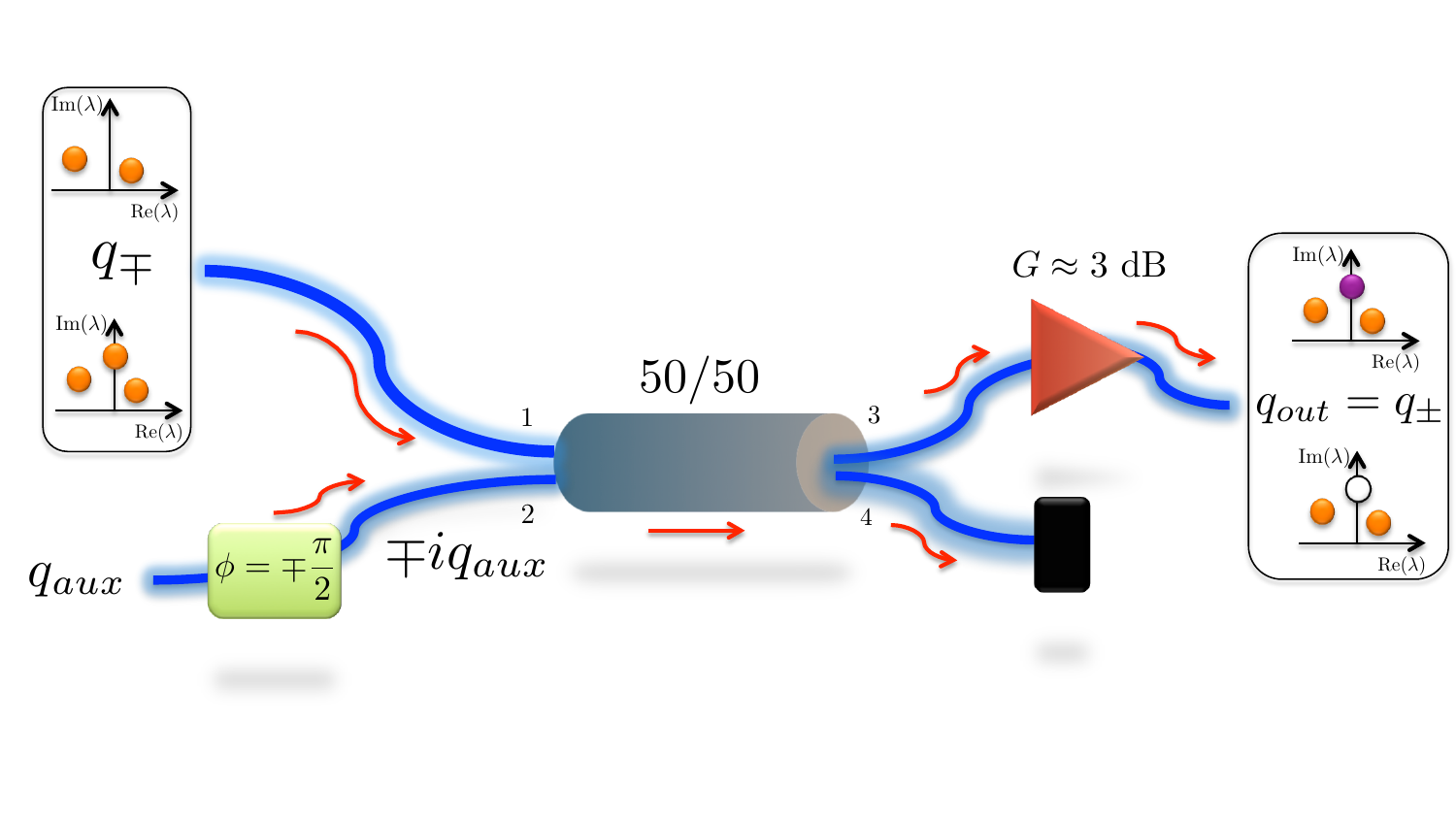}
    \caption{\emph{Schematic of the Optical Darboux Transformer}. In the multiplexer mode the signal $q_-$ (with spectrum featuring 2 discrete eigenvalues) is combined at a 50/50 coupler with the auxiliary signal $q_{aux}$ phase shifted by $-\pi/2$. In the filter mode the input signal is instead $q_+$, featuring three spectral eigenvalues, and is combined with $q_{aux}$ phase-shifted by $\pi/2$. The output from port 3 is amplified with 3 dB gain by an optical amplifier (red triangle) and results in signal $q_{out}$ featuring a new eigenvalue with respect to $q_-$ (in purple), or has one eigenvalue removed with respect to $q_+$ (white circle) depending on the filter or multiplexing operational mode.}
    \label{fig:1}
\end{figure}
Based on the known formalism for the iterative addition of individual eigenvalues we can hence proceed to define the explicit expression for the DT associated to the addition of \emph{multiple} eigenvalues to a signal $q_{-}$ by coherently combining it with a suitable \emph{single} auxiliary signal $q_{aux}$. We notice that we can always write Eq.\ref{DT2} as 
\begin{eqnarray}
q_+=q_{-}+q_{aux}
\end{eqnarray}
where $q_-$ is a signal featuring a certain  continuous spectrum $Q_{-}(\lambda)$ and a discrete spectrum with $\{\lambda_1,...,\lambda_m\}$ eigenvalues and spectral amplitudes $\{b_1,...,b_m\}$; while $q_{aux}$ is an auxiliary signal such that $q_+$ features
 $\{\lambda_1,...,\lambda_{m},\lambda_{m+1},..,\lambda_n\}$
eigenvalues with spectral amplitudes $\{b_1,...,b_m,b_{m+1},...,b_n\}$. 
In particular 
\begin{eqnarray}
q_{aux}=\sum_{i=m+1}^n\mathcal{D}_i
\end{eqnarray}
provided that the functions $v_{1,2}^{(k)}$ ($k=m+1,...,n$) -- which determine $\mathcal{D}_i$ -- have been calculated according to Eq.\ref{vk} keeping into account that the transformation operated on them is a function \emph{also} of the eignevalues of $q_{-}$, $\{\lambda_1,...,\lambda_m\}$.
It follows that adding eigenvalues $\{\lambda_{m+1},..,\lambda_{n}\}$ to the spectrum of $q_{-}$ requires simply coherently combining together the waveforms $q_{-}$ and $q_{aux}$. We can hence consider a device, schematically illustrated in Fig.\ref{fig:1}, that we call the \emph{Optical Darboux Transformer} (ODT) for solitons, which can be built using a 50/50 optical coupler, a phase shifter and an optical amplifier. 
In the eigenvalue multiplexing operational regime, the input at port 1 of the coupler is $q_{-}$ while the input at port 2 is $q_{aux}$ after being phase shifted by $-\pi/2$ radians. In this way the output of port 3 reads $q_3=\frac{1}{\sqrt{2}}q_{-}+i\frac{1}{\sqrt{2}}(-iq_{aux})$ and after amplification  by a factor of $2$ for power ($\sqrt{2}$ for amplitude) corresponding to $\approx 3$ dB gain, results indeed in $q_{out}=\sqrt{2}q_3=q_+$ which has eigenvalues $\{\lambda_1,..,\lambda_n\}$. It is easy to notice that the ODT can operate as an NFT filter too, which removes eigenvalues $\{\lambda_{m+1},...,\lambda_{n}\}$ from the spectrum $\{\lambda_1,...,\lambda_{n}\}$ of the signal $q_+$. In that case we call $q_+$ the input signal entering from port 1 of the 50/50 coupler while the phase shifter is set to provide a $+\pi/2$ radians phase shift to $q_{aux}$. The output of port 3 after the amplifier in this case reads $q_3=\sqrt{2}\left[\frac{1}{\sqrt{2}}q_++i\frac{1}{\sqrt{2}}(iq_{aux})\right]=q_{-}$, 
which has discrete spectrum $\{\lambda_{1},...,\lambda_{m}\}$.
Interestingly, the possibility of using the DT to iteratively remove single eigenvalues form the nonlinear discrete spectrum has been suggested mathematically\cite{Lin} and explored for developing efficient algorithms for NFT based optical communication systems digital data processing\cite{Aref2}.
We also stress the fact that the operations described above, both concerning the multiplexing and the filtering of eigenvalues, are \emph{not} equivalent to simply coherently adding or subtracting, to a given signal, an auxiliary signal containing hyperbolic secant waveforms having the eigenvalues one wishes to add or remove to/from the signal nonlinear spectrum. As it can be appreciated from examples reported below, $q_{aux}$ exhibits in general a much richer spectrum. Indeed the NFT operation $\mathcal{N}[\cdot]$ is \emph{nonlinear}, namely in general $\mathcal{N}[q_1]+\mathcal{N}[q_2]\ne\mathcal{N}[q_1+q_2]$. 
This nonlinearity is also a reason why an amplifier is needed at the coupler port 3 output, indeed, unless the signal power is very low $\mathcal{N}[a q_1]\ne a\mathcal{N}[q_1]$ with $a$ a complex constant.
In Figs. \ref{fig:2}a)-c) an example is shown of the ODT working as a multiplexer and performing $\emph{de facto}$ the function of a single soliton creation operator which adds one eigenvalue to the input signal spectrum.
In Figs. \ref{fig:2}d)-f) instead an example is shown of the ODT acting as a filter, by performing the role of a single soliton annihilation operator, and removing an eigenvalue from the input signal spectrum.
In both cases input, output and auxiliary signals and respective discrete eigenvalue spectrum are shown. It is interesting to notice that the auxiliary signal $q_{aux}$ features, besides a solitonic spectrum, a non vanishing continuous spectrum too, while $q_{\pm}$ don't. 
In Fig. \ref{fig:3} it is furthermore shown how the ODT can simultaneously add or remove multiple eigenvalues to/from the spectrum. Finally in Fig.\ref{fig:4} an example is shown where the ODT adds or removes an eigenvalue to/from an input signal  featuring a non vanishing continuous spectrum too\cite{note}. 
\newline
The simple architecture of the proposed device can be realised using existing commercially available optical fiber technology components like couplers, phase shifters and amplifiers such as Erbium-doped fiber ones. The auxiliary signal $q_{aux}$ can be synthesized using a waveform generator, a laser, and a modulator, while its necessary temporal synchronisation with the input waveform could be achieved for instance exploiting techniques used for signal synchronisation in time-division multiplexing for optical communications\cite{Weber}. In case the input signal to be processed is not known \emph{a priori}, information about its nonlinear spectrum can be obtained from detecting and analysing a small portion of it before it enters the ODT, in order to  synthesize a suitable $q_{aux}$ for the desired multiplexing or filtering operation.
The ODT presented in this work can find natural generalisation to the dual polarisation case (Manakov system) too. This can be implemented for instance using a first polarising beam splitter directing each polarization to a different ODT, and a second one for recombining the two polarizations after solitonic waveform processing (additional amplifiers could be employed where needed to compensate for power losses). 
\newline
It is important to stress a vital point. A distinctive conceptual feature of soliton filtering and soliton multiplexing in nonlinear Fourier domain is indeed the necessity of embedding a knowledge of the signal to be filtered/multiplexed in the filter/multiplexer itself. This is at variance with the operational principle of some traditional linear optical components such as standard optical filters which do not require any information about the input signal.
Similar paradigm changes, despite certainly appearing surprising for a way of thinking that is used to conceptualise the world based on a linear approach, are required when developing a truly nonlinear scientific \emph{Weltanschauung} -- worldview --, and a nonlinear technology too.
\newline
The ODT may find several applications in optical soliton technology. The possibility of optically controlling and manipulating solitonic waveforms, including adding and removing solitons from an optical signal or separating the continuous from the discrete spectrum, or filtering noise, are only some of the simplest and most obvious functionalities within reach. Among its broader applications the ODT could furthermore enable manipulation of the statistical properties of an optical soliton gas \cite{El1,El2,Gelash,Suret} both through its filtering and multiplexing functionalities. It could provide as well a novel tool for NFT-based analysis and control of the solitonic content of optical waves generated by lasers \cite{NFTlaser,Chekhovskoy}, both as an intracavity component and as an out-of-cavity optical signal processing tool too.
The proposed ODT, due to its potential cascadability and scalability could constitute an elementary component in more complex nonlinear Fourier domain signal processing and manipulation devices. To conclude, we have demonstrated the possibility of an optical domain implementation of the DT for the NLSE by means of a simple photonic architecture. The ODT offers control over solitonic waveforms by enabling selective discrete eigenvalues multiplexing and filtering. Further to that, by reducing the need of intermediate electrical and digital domain operations, the proposed device provides an advance towards optical signal processing in nonlinear Fourier domain.
\newline\newline
The author acknowledges support by the Royal Academy of Engineering through the Research Fellowship Scheme, and is grateful to Dr. Morteza Kamalian-Kopae and Dr. Mingming Tan for useful discussions.


\bibliography{apssamp}

\end{document}